\documentclass[conference]{IEEEtran}
\IEEEoverridecommandlockouts
\usepackage{cite}
\usepackage{amsmath,amssymb,amsfonts}
\usepackage{algorithm}
\usepackage{algpseudocode}
\usepackage{url}
\usepackage[pdftex]{graphicx}
\usepackage{textcomp}
\usepackage{booktabs,ragged2e}
\usepackage[pdftex]{xcolor}

\newcommand{\etal}{{\it et al.}}

\def\BibTeX{{\rm B\kern-.05em{\sc i\kern-.025em b}\kern-.08em
    T\kern-.1667em\lower.7ex\hbox{E}\kern-.125emX}}
\begin{document}

\title{Faster Lead Optimization Mapper Algorithm for Large-Scale Relative Free Energy Perturbation}

\author{
\IEEEauthorblockN{Kairi Furui}
\IEEEauthorblockA{\textit{School of Computing} \\
\textit{Tokyo Institute of Technology}\\
Kanagawa, Japan \\
furui@li.c.titech.ac.jp}
\and
\IEEEauthorblockN{Masahito Ohue}
\IEEEauthorblockA{\textit{School of Computing} \\
\textit{Tokyo Institute of Technology}\\
Kanagawa, Japan \\
ohue@c.titech.ac.jp}
}

\maketitle

\begin{abstract}
In recent years, free energy perturbation (FEP) calculations have garnered increasing attention as tools to support drug discovery.
The lead optimization mapper (Lomap) was proposed as an algorithm to calculate the relative free energy between ligands efficiently.
However, Lomap requires checking whether each edge in the FEP graph is removable, which necessitates checking the constraints for all edges.
Consequently, conventional Lomap requires significant computation time, at least several hours for cases involving hundreds of compounds, and is impractical for cases with more than tens of thousands of edges.
In this study, we aimed to reduce the computational cost of Lomap to enable the construction of FEP graphs for hundreds of compounds.
We can reduce the overall number of constraint checks required from an amount dependent on the number of edges to one dependent on the number of nodes by using the chunk check process to check the constraints for as many edges as possible simultaneously.
Moreover, the output graph is equivalent to that obtained using conventional Lomap, enabling direct replacement of the original Lomap with our method.
With our improvement, the execution was tens to hundreds of times faster than that of the original Lomap.

\end{abstract}

{\small \it Regular Research Paper}

\section{Introduction}
In the early stages of drug discovery, hit-to-lead optimization is conducted to identify lead compounds with sufficient affinity and desirable pharmacological properties from hit compounds that bind weakly to the target receptor~\cite{bleicher2003hit}.
The alchemical free energy perturbation (FEP) calculation~\cite{wang2015accurate,kuhn2020assessment,gapsys2020large} is a popular tool for lead optimization because it uses a perturbation-based approach to calculate the exact relative binding free energies between a candidate compound and its target receptor from molecular dynamics simulations~\cite{muegge2023recent,schindler2020large}.

\begin{figure*}[tb]
    \centering
    \includegraphics[width=0.95\linewidth]{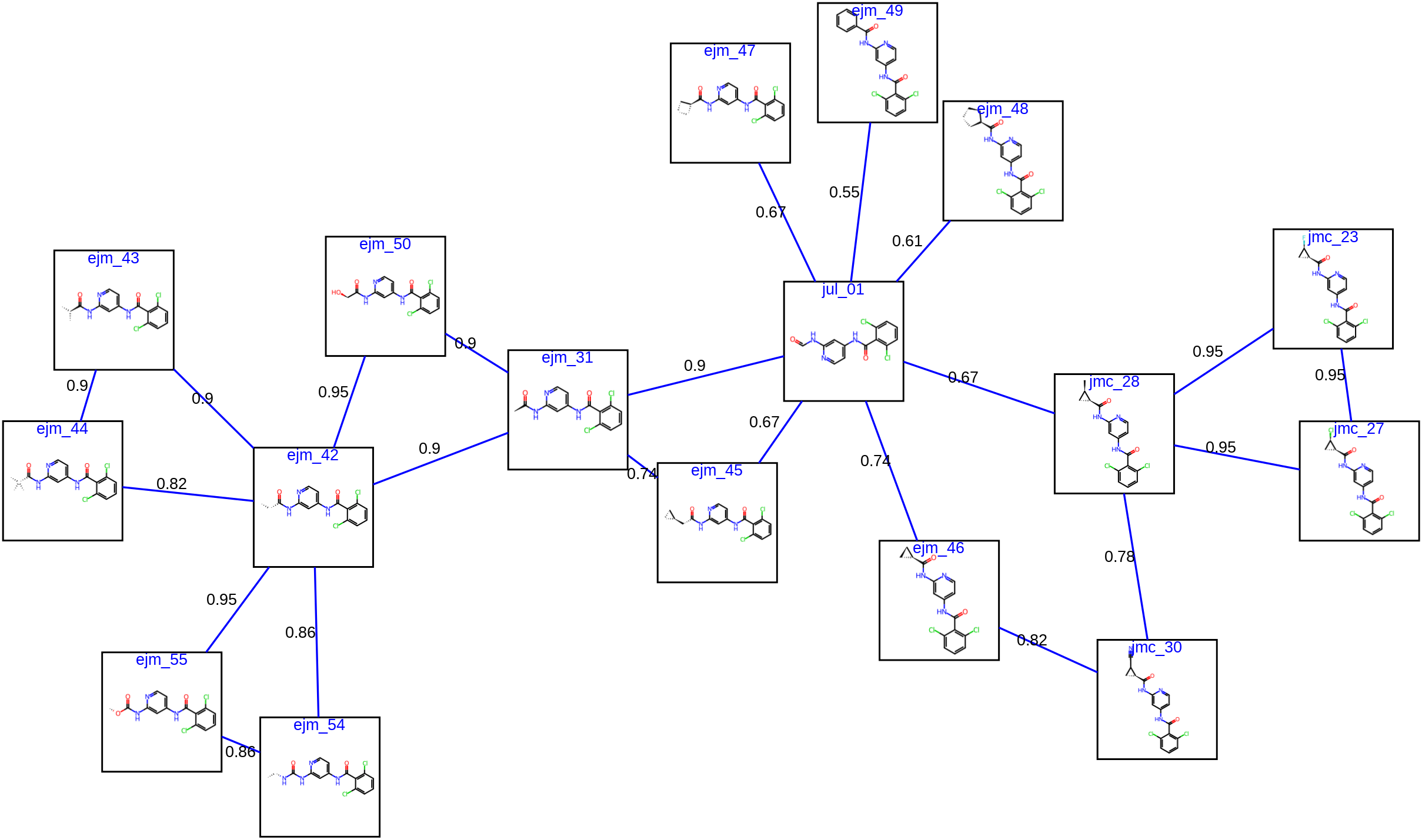}
    \caption{Example FEP graphs on the Wang-benchmark dataset (target:TYK2). Drawn using Lomap and NetworkX library.}
	\label{fig:example}
\end{figure*}

Closed thermodynamic cycles are crucial for minimizing errors in the relative binding free energy calculations~\cite{liu2013lead,xu2019optimal,pitman2022design}.
The sum of the relative free energies of the thermodynamically closed paths is theoretically zero. The error in a closed cycle is called the cycle closure error or hysteresis~\cite{liu2013lead}.
Cycle closure methods, which reduce errors based on closed cycles, can provide better prediction accuracy than perturbation calculations.
Therefore, to guarantee the accuracy of the FEP calculations, the FEP graph must include as many closed cycles as computational resources allow.
Fig.~\ref{fig:example} illustrates an example FEP graph of a TYK2 target benchmark series by Wang \etal~\cite{wang2015accurate} run with Cresset Flare's FEP application~\cite{bauer2019electrostatic,kuhn2020assessment}.

First, the most common FEP graph is the star graph, which connects all compounds to only one reference compound.
No particular graph construction algorithm is required, and the perturbation calculation is only performed $N-1$ times.
However, the accuracy of the FEP calculation cannot be guaranteed because no single closed thermodynamic cycle exists.
PyAutoFEP~\cite{carvalho2021pyautofep}, an open-source tool for automating FEP calculations provides an implementation that builds wheel-shaped and optimal graphs using colony optimization.

DiffNet~\cite{xu2019optimal} provides a theoretical paradigm for optimizing the covariance matrix for errors.
HiMap~\cite{pitman2022design} also provides a statistically optimal design based on optimization of the covariance matrix, similar to DiffNet.

Lomap~\cite{liu2013lead} is an automated planning algorithm for calculating relative free energy.
Lomap calculates the scores between compound pairs based on several design criteria for proper relative free energy calculations, and FEP graphs are constructed based on these scores.
Here, Lomap constructs an FEP graph based on the criterion that all molecules must be part of at least one closed cycle.
However, Lomap only handles the generation of FEP graphs for a few dozen compounds and must consider hundreds of FEP calculations.

Recently, active learning~\cite{settles2009active,reker2015active,reker2017active}, a machine learning approach for planning experiments for lead optimization, has been applied to relative binding free energy calculations~\cite{konze2019reaction,gusev2023active,khalak2022chemical}.
Active learning involves hundreds or thousands of extensive FEP calculations.
However, conventional Lomap calculations incur a substantial computational cost for FEP graphs of several hundred compounds.
Therefore, applying the Lomap graph directly to active learning cases requires significant effort.
Conventional active learning methods with FEP only attempt to perform star-shaped FEP calculations without closed cycles~\cite{gusev2023active,khalak2022chemical} or perform more detailed calculations for a limited number of compounds~\cite{konze2019reaction}.

Therefore, simplifying the introduction of FEP cycles, even for hundreds of compounds, is imperative to support FEP calculations by active learning.

The main reason why Lomap calculations require a considerable amount of time is that they check the constraints for every edge in the initial graph.

\begin{figure}[tb]
    \centering
    \includegraphics[width=\linewidth]{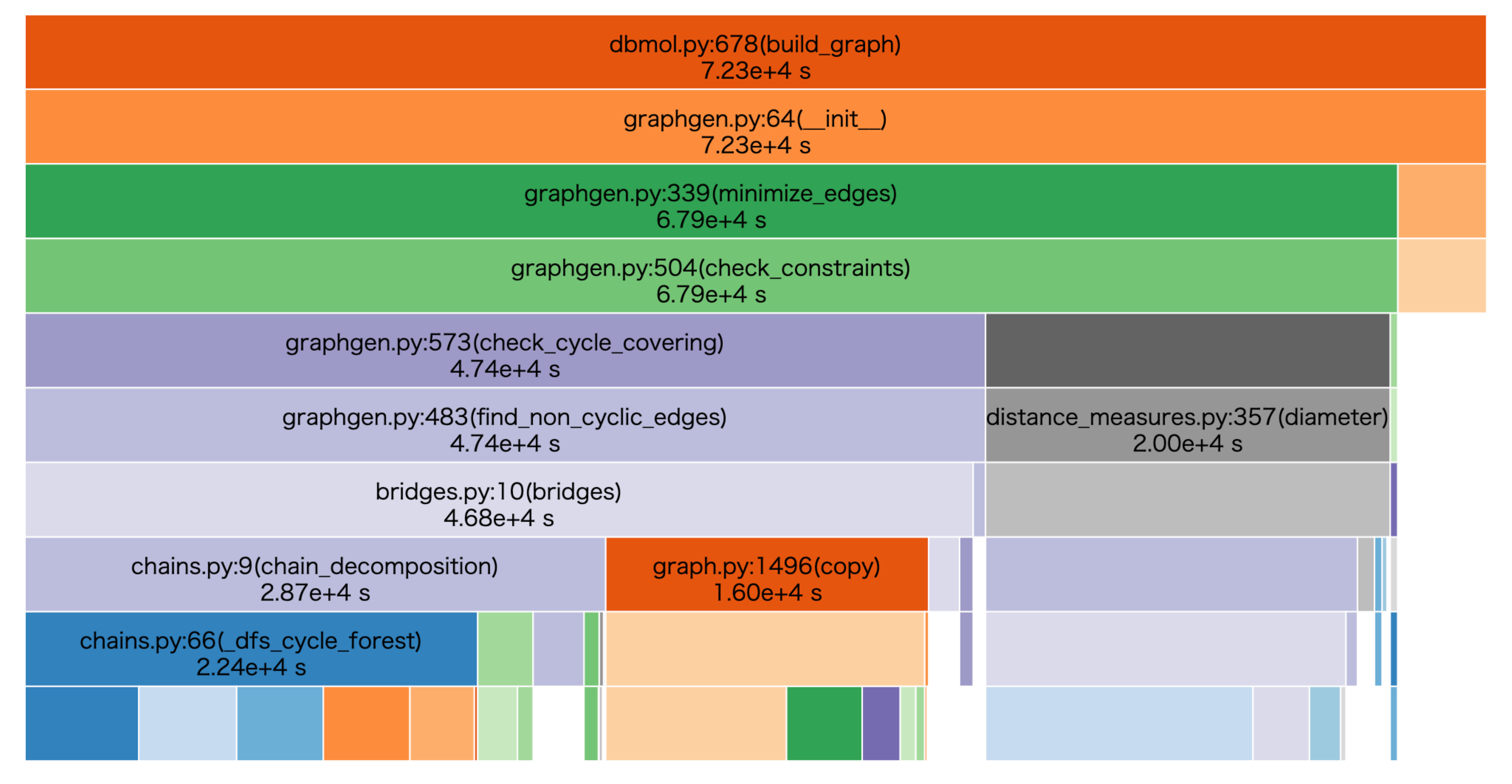}
    \caption{Results of profiles by cProfile of the baseline Lomap algorithm (Thompson TYK2 dataset, $N=1000$). Drawing by snakeviz.}
	\label{fig:profile}
\end{figure}

Fig.~\ref{fig:profile} shows the results of FEP graph generation using the Lomap algorithm for 1000 nodes.
Most of the execution time of the \texttt{build\_graph} function was occupied by the \texttt{check\_constraints} function.

In this paper, we propose an improved Lomap algorithm that introduces a chunk check process in which edges are processed collectively as much as possible to maintain the number of constraint checks dependent on the number of nodes.
The proposed method can construct FEP graphs tens to hundreds of times faster for several hundred compounds.
Furthermore, the output FEP graph was equivalent to that of the original Lomap.
\section{Methods}
\subsection{Lomap algorithm}
Lomap is available as an open-source implementation using RDKit (\url{https://github.com/OpenFreeEnergy/Lomap}), and this study is based on and refined from it.
In this section, we describe the algorithm used in the original Lomap.
Lomap generates an undirected graph that determines the compound pairs for which relative binding free energy calculations will be conducted.
The FEP graph must consist of edges for atom-to-atom pairs that are sufficiently similar to calculate valid free energies.
The link score is an index based on the similarity between two compounds.
The link score is calculated based on the maximum common substructure (MCS) of a compound pair using the following equation:
\begin{align}
    S=\exp[-\beta (N_A+N_B-2N_{MCS})],
\end{align}
where $\beta$ is a parameter, and $N_A$, $N_B$, and $N_{MCS}$ are the number of heavy atoms in the two input compounds and MCS, respectively.
That is the number of insertions and deletions required for the perturbation.
The score was defined between 0 and 1, with a maximum of 1 when the two compounds were the same.
The score includes a correction that takes a product of other scores calculated based on whether they are valid for the perturbation calculation, such as having the same net charge and preserving the rings as much as possible.
A crucial aspect of graph generation is that edges with low link scores should be removed except when they favorably impact a closed cycle.
Computing the link scores among all the compounds is required to generate an FEP graph using the Lomap algorithm.
In this study, the score calculation for each compound pair was not subject to acceleration because it could be parallelized.
In addition, score calculation is not subject to execution time measurement because the scores are calculated in advance.

Lomap's FEP graph generation is explained as follows:
\begin{enumerate}
    \item Compute a scoring matrix for all compound-to-compound pairs.
    \item The initial graph should consist of only the edges of pairs whose link scores are above the threshold. The initial graph may be partitioned into multiple subgraphs.
    \item The edge removal process is executed for each connected component of the initial graph. First, sort the edges in the subgraph in order of descending link score.
    \item \label{lomap:process4} Begin with the edge having the lowest link score and assess whether the constraints are met upon its removal. If the constraints are satisfied, remove the edge; otherwise, leave it in place. The four constraints are as follows:
    \begin{enumerate}
        \item The subgraph is connected.
	\item The number of nodes not included in the cycles does not change. This constraint can be determined by the number of bridges in the cycle not changing.
	\item The diameter of the subgraph is less than or equal to the threshold MAXDIST.
	\item (if necessary) The distance from the known active ligand is less than or equal to the threshold.
    \end{enumerate}
    \item Connect each connected component.
\end{enumerate}

Here, Process (\ref{lomap:process4}) required the most time when the graph was large.
This is because Process (\ref{lomap:process4}) checks all edges in the subgraph to determine whether an edge can be removed.
The graph diameter and bridges were then calculated to determine whether an edge could be removed.
If the number of nodes in the graph is $N$ and the number of edges is $M$, the computational complexity of the diameter is $O(MN)$, and that of the bridges is $O(M+N)$.
The computational complexity becomes at least $O(M^2N)$ because the number of edges in the graph changes each time the FEP graph is updated.
The most significant factor in this computational complexity is that it checks all edges to determine whether they can be removed.
The Lomap algorithm removes edges in the order of the worst score; however, this order must be preserved.
Therefore, the graph-generation phase cannot be computed in parallel.

\subsection{Chunk Check Process}
\begin{figure*}[tb]
    \centering
    \includegraphics[width=0.9\linewidth]{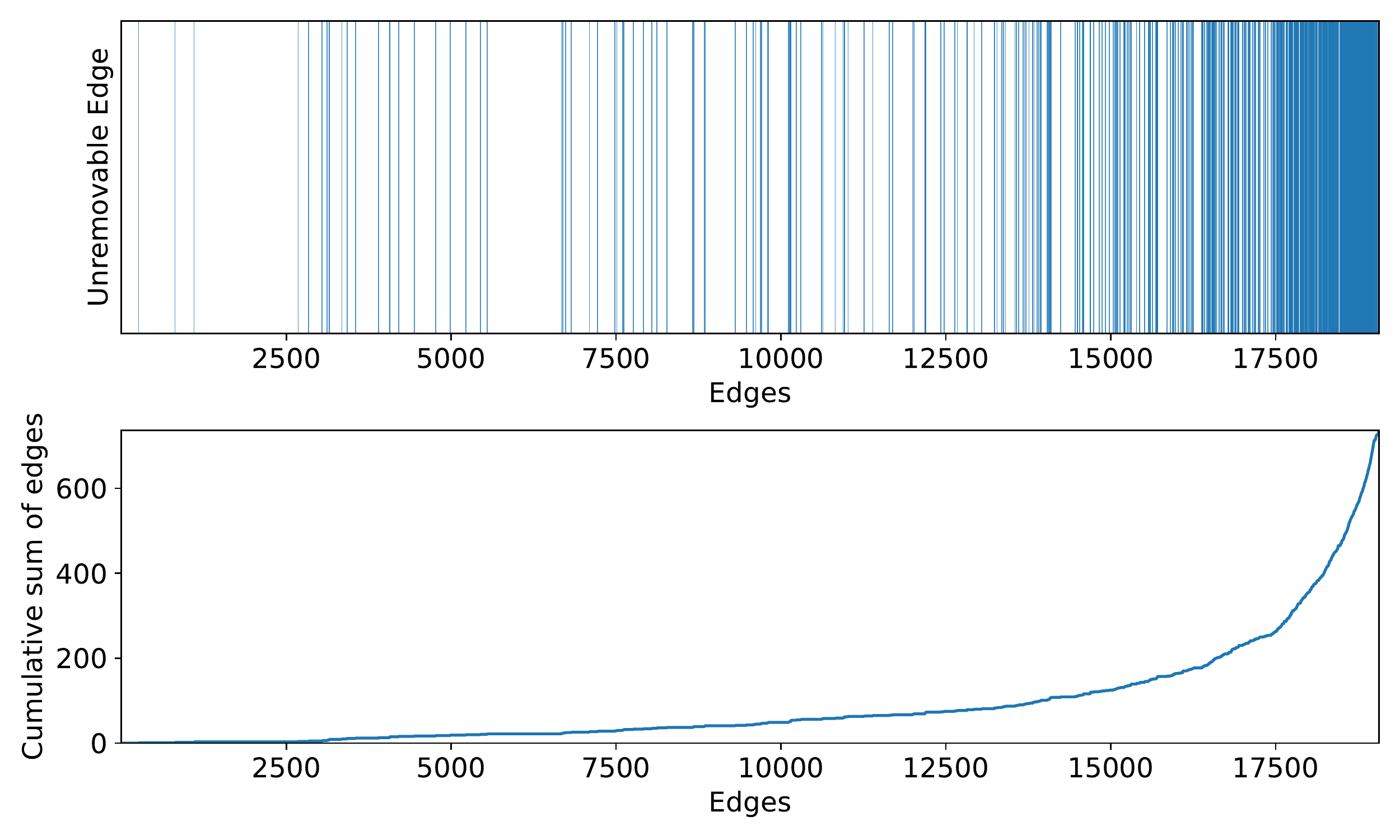}
    \caption{Occurrence of unremovable edges and their cumulative sum across constraint checks (Thompson TYK2 dataset, 500 nodes). Among all possible edge combinations, 19,066 edges with a link score of 0.4 or higher are checked, resulting in 740 edges for the outcome.}
    	\label{fig:edges_500}
\end{figure*}

In this section, we propose the chunk check process.
Because the number of edges in the initial graph is proportional to the square of the number of nodes, if the link score is not pruned with a threshold, the ratio of edges that do not satisfy the constraint follows the inverse of the number of nodes.
In other words, the larger the number of nodes, the greater the number of edges that can be removed.
Edges that do not satisfy the constraints are unevenly distributed toward the end of the checking process.
Fig.~\ref{fig:edges_500} illustrates this tendency during the edge removal process.
This is because numerous redundant edges exist at the outset of the process, making it easier to satisfy the constraints by removing the edges. However, as the procedure progressed, the conditions became more stringent, making edge removal more challenging.
Therefore, these blank areas should be removed as much as possible in a single check, and the edges need to be removed individually in areas with a concentration of unremovable edges.

\begin{figure}[tb]
    \centering
    \includegraphics[width=\linewidth]{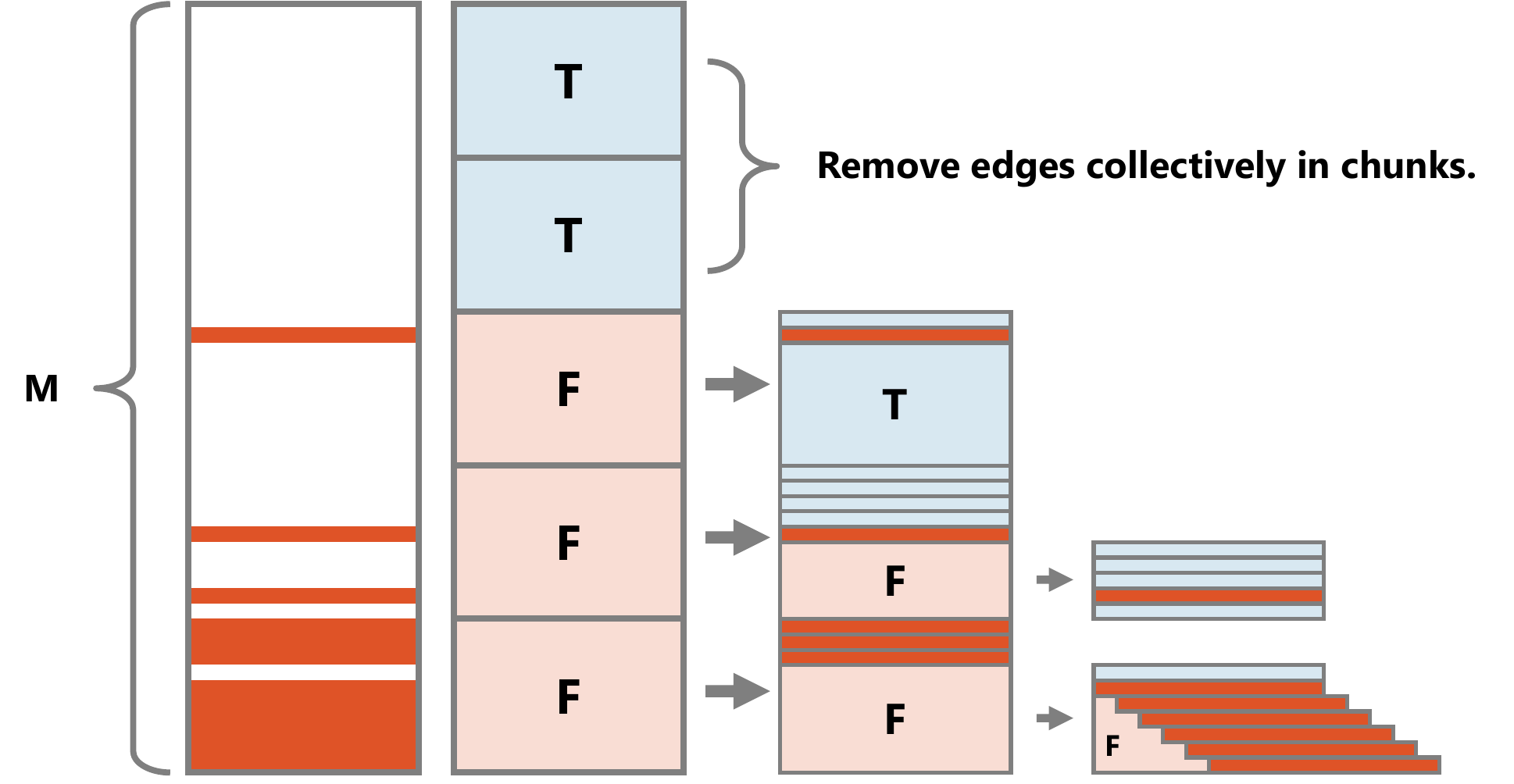}
    \caption{Overview of Chunk Check Processing.}
	\label{fig:chunk_check}
\end{figure}

Given these observations, we propose a chunk check process.
Fig.~\ref{fig:chunk_check} provides an overview of the chunk check process.

First, let the chunk scale $s$ be a parameter related to the size of the chunk, let the chunk level $k$ be the depth of the layer of the chunk process, and let $C_{k}$ be the chunk when chunk level $k$ is set.
In addition, for the number of edges $M$, $k_{max}=\lfloor\log_{s}M \rfloor$ represents the maximum chunk level, indicating the number of times that chunk processing can be executed recursively.
The chunk size for the first $k=0$ was $S_0=s^k_{max}$.
For example, when $M=5100$ and $s=10$, $S_0=s^{k_{max}}=10^3$.
The algorithmic procedure for the chunk check process is as follows:
\begin{enumerate}
    \item \label{enu:step1} Check the constraints of step (\ref{enu:step2}) for every $S_0$ edge sorted in order of decreasing link score.

    \item \label{enu:step2} Constraint check.
    If the constraint is not satisfied (i.e., there are some unremovable edges in the chunk), the check fails, and all edges are restored and go to the next step (\ref{enu:step3}).
    The constraint check succeeds when the constraint is satisfied even if $S_0$ edges are removed.
    If the constraint check succeeds for all edges, return to step (\ref{enu:step1}).
    \item \label{enu:step3} As $S_{k}=\max(\lfloor S_{k-1}/s\rfloor , 1)$, check $C_k$ constraints for each $S_{k}$ edges in chunk $C_{k-1}$.
    \item \label{enu:step4} If the undeletable edges are identified, check the constraints for the remaining edges of chunk $C_{k-1}$. If the constraint is satisfied, complete the chunk process for chunk $C_{k-1}$ and return to step (\ref{enu:step2}). If not satisfied, continue with step (\ref{enu:step3}) until the chunk check of all edges in the chunk succeeds.
\end{enumerate}

The chunk check process allows the removal of each large chunk in areas with few unremovable edges.
When the chunk scale is large, unremovable edges can be identified quickly in areas with numerous unremovable edges.

\subsection{Minor modifications to diameter calculations}
The original Lomap implementation used the NetworkX~\cite{SciPyProceedings_11} eccentricity method, but we changed it to calculate the diameter using the bounding diameter algorithm~\cite{takes2011determining,borassi2015fast}.
This algorithm is available in the NetworkX diameter function with the \texttt{usebounds} option.
The diameter calculation is usually $O(MN)$ computation complex because it computes the all-pairs shortest path (ASAP).
The algorithm uses the lower and upper boundary relationships between node eccentricity and diameter to determine the exact diameter within a computation time proportional to $O(N)$.
Border cases, including complete graphs and circular graphs, should be considered, but should also be considered when constructing FEP graphs.

\subsection{Code availability}
Our proposed improved version of the Lomap code (called FastLomap) is available at \url{https://github.com/ohuelab/FastLomap} under the same MIT license as the original Lomap.

\section{Experiments}
\subsection{Thompson TYK2 dataset}

In this study, we used 100–1000 nodes for extensive FEP graph construction experiments.
Therefore, we used a congeneric series of TYK2 inhibitors based on the aminopyrimidine scaffold provided by Thompson \etal~\cite{thompson2022optimizing}.

\begin{figure}[t]
    \centering
    \includegraphics[width=0.95\linewidth]{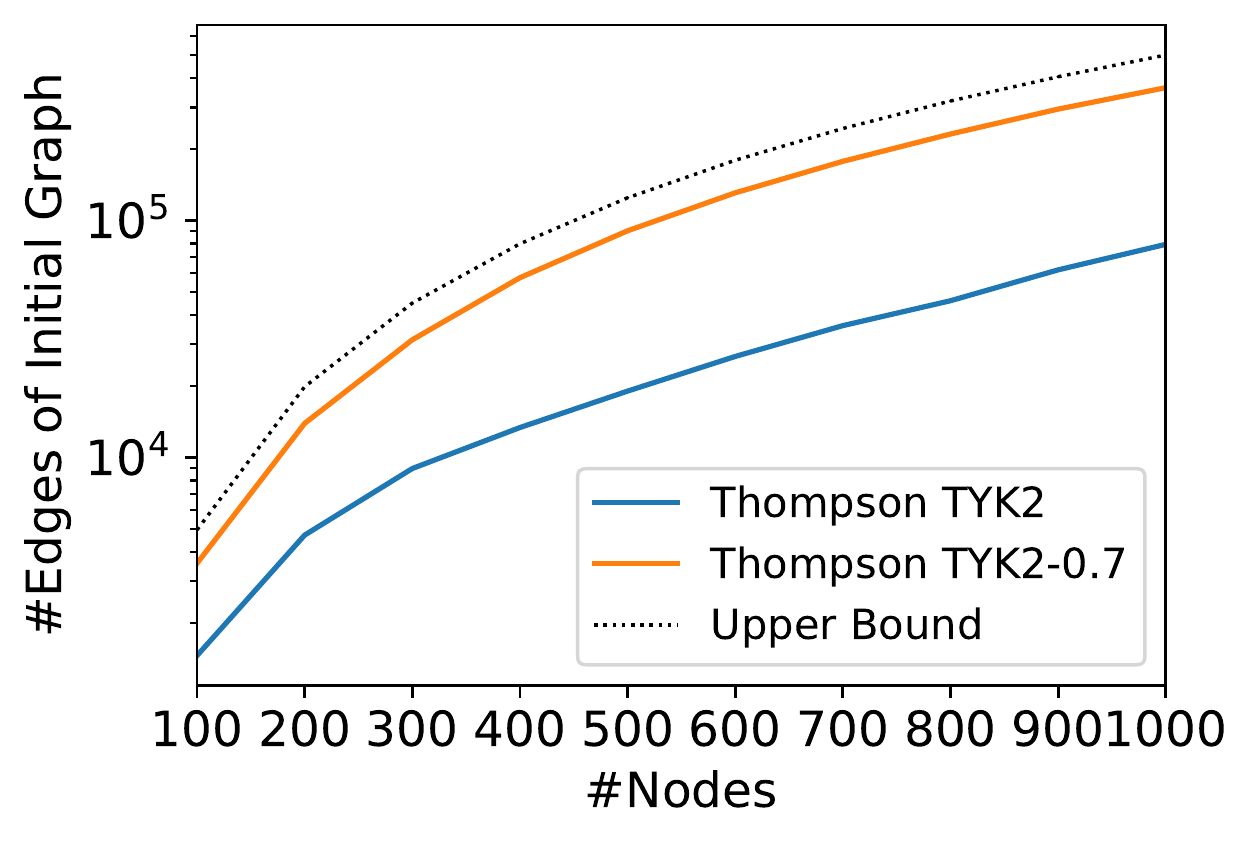}
    \caption{ The number of edges relative to the number of nodes of the initial graph in Thompson TYK2 datasets.}
	\label{fig:num_edges}
\end{figure}

\begin{figure}[t]
    \centering
    \includegraphics[width=0.99\linewidth]{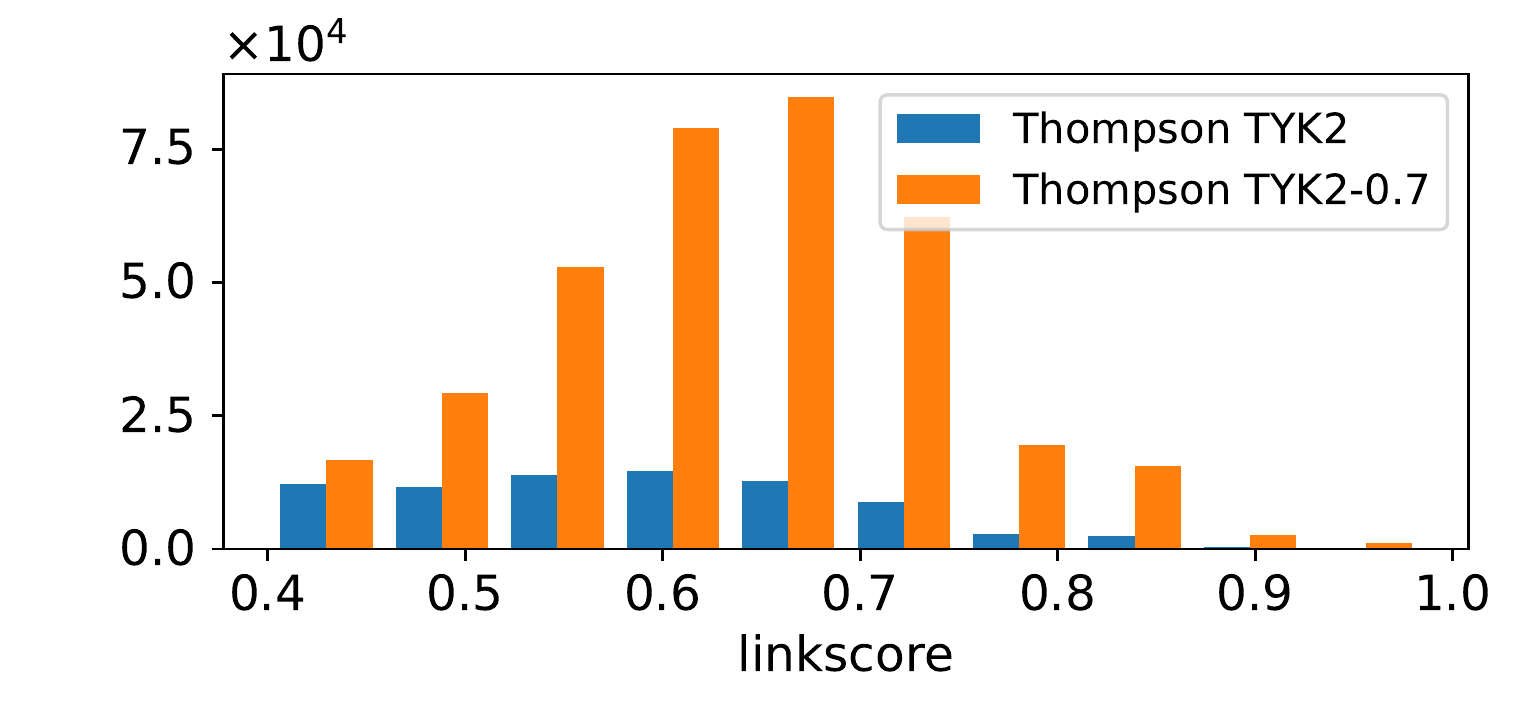}
    \caption{Histogram of link scores for Thompson TYK2 dataset.}
	\label{fig:hist_linkscore}
\end{figure}

\begin{figure}[t]
    \centering
    \includegraphics[width=0.95\linewidth]{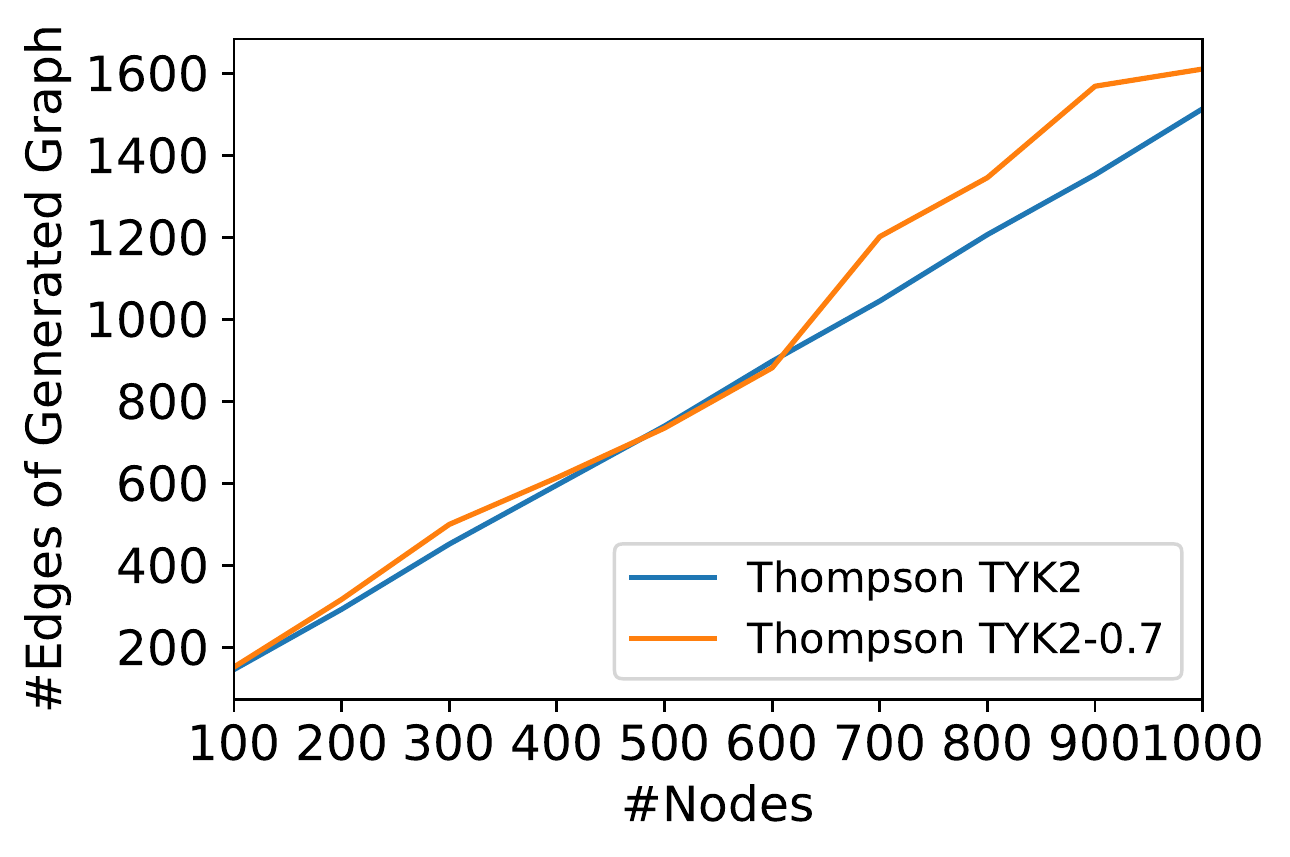}
    \caption{ The number of edges relative to the number of nodes of the generated graph in Thompson TYK2 datasets.}
	\label{fig:num_edges_opt}
\end{figure}

        \begin{description}
    \item[(a) Thompson TYK2 dataset]\mbox{}\\ From the original dataset, we randomly selected 1000 cases and constructed a subset of the experimental set from 100 to 1000 nodes.
    \item[(b) Thompson TYK2$_{0.7}$ dataset]\mbox{}\\ From the original dataset with a link score of 0.7 or higher with the reference ligand (PDB: 4GIH~\cite{liang2013lead}), the experimental set was constructed from 100 to 1000 cases in the same way.
\end{description}
Fig.~\ref{fig:num_edges} illustrates the number of edges in the initial graph against the number of nodes for the Thompson TYK2 and Thompson TYK2$_{0.7}$ datasets.
The edges of the initial graph were pruned to a link score threshold of 0.4.
Therefore, the number of edges in the Thompson TYK2$_{0.7}$ dataset, which consisted of similar compounds, was at most five times larger than that in the Thompson TYK2 dataset.
Fig.~\ref{fig:hist_linkscore} shows the distribution of link scores for the two datasets.
The peak of the link score for the Thompson TYK2$_{0.7}$ dataset was higher than that for the Thompson TYK2 dataset.
The numbers of edges in the generated FEP graph for each node are shown in Fig.~\ref{fig:num_edges_opt}.
Both datasets have a similar number of edges in the FEP graph.
Note that the generated graphs are equivalent between the proposed method and the existing Lomap.

\subsection{Wang dataset}
The Wang dataset~\cite{wang2015accurate} was used as a benchmark to evaluate RBFE prediction performance and contained several dozen ligands for each of the eight targets.
Our proposed algorithm assumes that the number of compounds exceeds 100, and its benefit is small for such a dataset.
However, we used the Wang dataset to demonstrate that the proposed method can be applied to a general use case.

\subsection{Setup for graph generation}

First, we generated a complex structure because no complex structure was provided for the Thompson TYK2 dataset.
Complex structures were generated by the superposition of a reference ligand (PDB: 4GIH~\cite{liang2013lead}) using the alignment tools (\texttt{very-permissive} option) of Cresset Flare~\cite{bauer2019electrostatic}.

All link scores were computed using Lomap's default 3D mode.
Note that the link score calculation for all pairs of 1000 compounds takes 26.4 minutes in parallel processing on 28 CPUs.

We experimented with the original Lomap calculation as a baseline and with chunk scales $s=2, 5, 10, 50$ for the proposed method.
To generate the FEP graph, we used the default settings of Lomap, that is, a graph diameter of 6, no tree structure allowed, and a link score cutoff of 0.4.

\section{Results and Discussion}
\subsection{Results of Thompson TYK2 dataset}
\begin{figure*}[tb]
    \centering
    \includegraphics[width=0.99\linewidth]{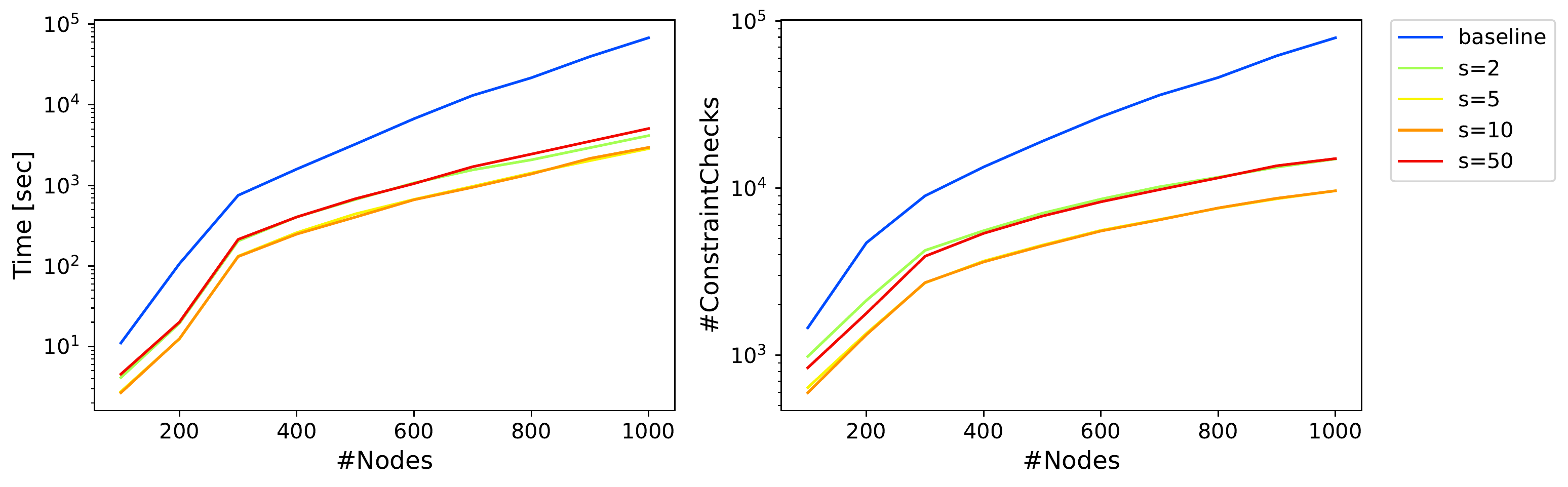}
    \caption{Execution time and number of constraint checks in the Thompson TYK2 dataset.}
	\label{fig:scale_results}
\end{figure*}

\begin{figure*}[tb]
    \centering
    \includegraphics[width=0.99\linewidth]{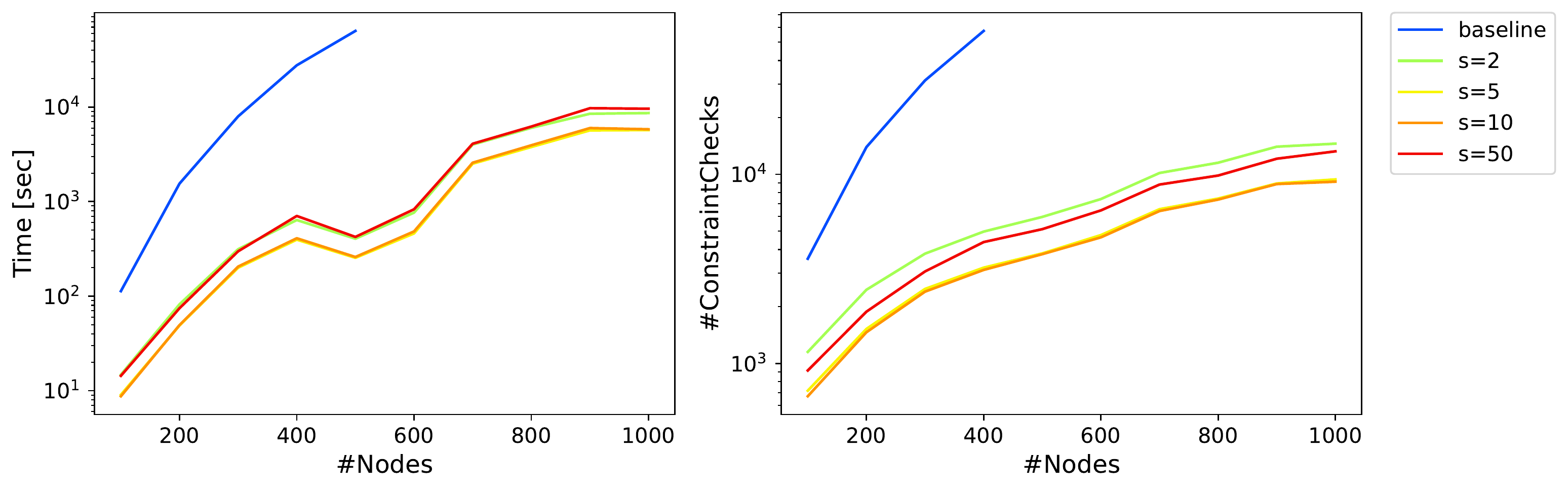}
    \caption{Execution time and number of constraint checks in the Thompson TYK2$_{0.7}$ dataset.}
	\label{fig:scale07_results}
\end{figure*}

Figs.~\ref{fig:scale_results} and \ref{fig:scale07_results} show the execution time and the number of constraint checks for each number of nodes in the Thompson TYK2 and Thompson TYK2$_{0.7}$ datasets.
The baseline method did not complete the calculation within 24 h for $N=500$ or more in the Thompson TYK2$_{0.7}$ dataset.
First, we compared the baseline and the proposed methods.
For the Thompson TYK dataset, it is approximately 23.7 times faster at most at $N=1000$.
In the case of the Thompson TYK2$_{0.7}$ dataset, the FEP graph was several hundred times faster at $N=400$, and the baseline method could not construct an FEP graph at $N=500$ or higher.
The Thompson TYK2$_{0.7}$ dataset has many edges in the initial graph owing to the poor pruning effect. Consequently, it is evident that the baseline method, which performs constraint checks an equal number of times as the number of edges, is unsuitable.
Hence, the proposed method is effective for cases involving many edges because the number of constraint checks depends on the number of nodes.

Next, the chunk scale score was evaluated.
For both datasets, the chunk scale was less for the execution time and the number of constraint checks when $s=5, 10$ than $s=2, 50$.
This was better than $s=2$ because a larger chunk scale reduced the maximum chunk level $k_{max}$ and promptly identified unremovable edges.
Conversely, if the chunk scale is larger, such as $s=50$, the chunk process becomes less effective in areas with few unremovable edges because a single chunk check failure leads to splitting into finer chunks.
Moreover, it is essential to note that the number of constraint checks and execution time do not directly correspond because deleting and restoring edges consumes time.
The number of constraint checks for $s=50$ is lower than that for $s=2$ in Fig. \ref{fig:scale07_results}, but the execution time is comparable due to the time to remove and restore edges.

\subsection{Results of Wang dataset}
\begin{figure*}[tb]
    \centering
    \includegraphics[width=0.99\linewidth]{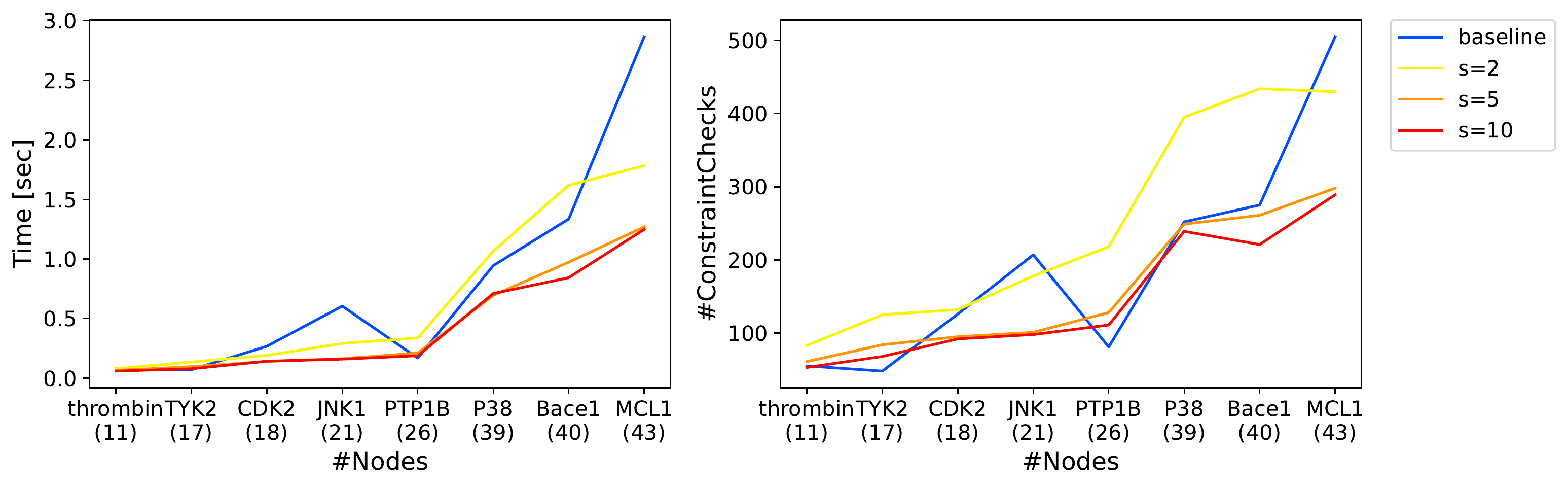}
    \caption{Execution time and number of constraint checks in the Wang dataset.}
	\label{fig:benchmark_results}
\end{figure*}

Fig.~\ref{fig:benchmark_results} exhibits the execution time and the number of constraint checks for each number of nodes in the Wang dataset.
Although the execution time increases with the number of nodes, even for $s=5, 10$, the execution time is slightly shorter than that of the baseline.
Even at baseline, the process took less than 3 s at most. Therefore, the proposed method has no negative impact on general use cases.

\section{Conclusion}
In this study, we propose a chunk check process to improve the Lomap algorithm and make it scalable for increasing the number of compounds.
The proposed method may have limited effectiveness for general use in a few dozen cases; however, it demonstrates significant efficiency and is several dozen times faster when applied to several hundred cases.
As the output graph of the proposed method is equivalent to that of the original Lomap algorithm, it can be used immediately to replace existing methods.

Although not addressed in this study, accelerating link score calculation is an issue for future research.
Because link score calculations for all compound pairs are required, the computational complexity increases by an order of squares.
The recently proposed HiMap~\cite{pitman2022design} uses the DBSCAN method for unsupervised clustering of similar compounds.
Clustering using compound fingerprints and calculating the scores based on the maximum common substructure for each cluster may reduce the overall computational cost.
In addition, further discussion is required on whether the Lomap's graphs is appropriate for large FEP designs.
It is crucial to analyze the impact of cycle–closure redundancy on the accuracy of perturbation maps in large FEP graphs.

Our proposed method can generate graphs up to 20 times faster than the original Lomap for cases with fewer edges and up to several hundred times faster for those with numerous edges.
The proposed method is not only applicable to the planning of large-scale FEP experiments but is also promising as an active learning approach for relative binding free energy calculations.

\section*{Acknowledgment}
This study was partially supported by JST FOREST (JPMJFR216J), JST ACT-X (JPMJAX20A3), JSPS KAKENHI (20H04280, 22K12258), and AMED BINDS (JP22ama121026).

\bibliographystyle{IEEEtran}
\bibliography{mainbib}

\end{document}